\pdfoutput=1
\documentclass[
reprint,
superscriptaddress,
frontmatterverbose,
amsmath,
amssymb,
aps,
prapplied,
nofootinbib,
longbibliography
]{revtex4-2}

\usepackage[french,english]{babel}
\usepackage{graphicx}
\usepackage{braket}
\usepackage{mathtools}
\usepackage{physics}
\usepackage{amsmath}
\usepackage{pdfpages}
\usepackage{pgffor}

\makeatletter
\AtBeginDocument{\let\LS@rot\@undefined}
\makeatother
\begin{document}

\title{Coexistence of broadband conduction and amplitude-driven topological confinement in synthetic acoustic matter}

\begin{abstract}
Nonlinear and topological wave phenomena offer new opportunities for robust wave manipulation. Existing experimental platforms have largely focused on on-site (Kerr-like) nonlinearities within tight-binding frameworks, leaving the regime of programmable nonlinear \emph{inter-site} coupling in far-field scattering systems essentially unexplored. Here we demonstrate a programmable acoustic metamaterial that exhibits amplitude-driven topological confinement of sound while simultaneously maintaining broadband linear transmission across the same frequency range. The system consists of an open acoustic waveguide coupled to a chain of actively controlled electroacoustic resonators enabling tunable nonlinear and nonlocal interactions. Using a classical wave-scattering model we show that increasing the excitation amplitude drives a robust topological interface mode localized deep within the structure, with the confinement strengthening continuously with drive rather than at a finite threshold. The predictions are verified experimentally using a fully programmable metacrystal composed of individually controlled acoustic resonators, enabling real-time manipulation of coupling parameters. The mode's topological origin is established by symmetry: it survives coupling disorder beyond 50\% of nominal but collapses when a delay between adjacent couplings is introduced, effectively breaking the protecting chiral symmetry. These results establish a versatile platform for studying nonlinear topological wave physics and for designing adaptive acoustic metamaterials.
\end{abstract}

\author{Mathieu~Padlewski}
\author{Maxime~Volery}
\affiliation{Signal Processing Laboratory 2, EPFL, 1015 Lausanne, Switzerland}
\author{Romain~Fleury}
\affiliation{Laboratory of Wave Engineering, EPFL, 1015 Lausanne, Switzerland}
\author{Hervé~Lissek}
\author{Xinxin~Guo}
\affiliation{Signal Processing Laboratory 2, EPFL, 1015 Lausanne, Switzerland}

\maketitle

\section{Introduction}

Over the past decades, topology has emerged as a unifying framework across many areas of physics, providing compact geometric descriptions of wave phenomena with remarkable robustness to defects and parametric perturbations~\cite{moore_birth_2010}. This robustness is particularly advantageous in systems where backscattering limits performance, such as communication and energy transport networks.

To date, most realizations of topological phases have been confined to linear regimes. This prevalence reflects not only conceptual simplicity but also experimental practicality, as linear systems are more tractable to analyze and implement. In contrast, controlled nonlinearities enable a broad range of functionalities, including highly directive acoustic emission via parametric arrays~\cite{westervelt_parametric_1963}, vortex-based energy harvesting~\cite{le_fouest_optimal_2024}, and nonreciprocal transport~\cite{guo_observation_2023}. Extending topological concepts into nonlinear regimes therefore represents both a fundamental challenge and a promising route toward enhanced wave control. In parallel, the limited availability of naturally occurring topological platforms—particularly in quantum systems—has motivated the development of engineered analogues in metamaterials. These include nonlinear electrical circuits~\cite{hadad_self-induced_2018,zangeneh-nejad_nonlinear_2019,dobrykh_nonlinear_2018,wang_topologically_2019,serra-garcia_observation_2019}, nonlinear optical systems~\cite{smirnova_nonlinear_2020,maczewsky_nonlinearity-induced_2020,mukherjee_observation_2020,hu_nonlinear_2021}, mechanical structures~\cite{snee_edge_2019,lo_topology_2021,chaunsali_self-induced_2019,darabi_tunable_2019,chaunsali_stability_2021}, and active acoustic platforms that readily incorporate nonlinearity and non-Hermiticity~\cite{rivet_constant-pressure_2018,xia_nonlinear_2021,padlewski_active_2023,padlewski_observation_2025}. Despite this diversity, two dominant paradigms prevail. First, most studies focus on Kerr-like \emph{on-site} nonlinearities, which are favored for their passive implementation and their direct connection to the nonlinear Schrödinger equation~\cite{landau_theory_1950,ginzburg_macroscopic_1956}. The case of nonlinear \emph{inter-site} (coupling) nonlinearities — where the hopping amplitude itself depends on field intensity — is comparatively unexplored, and has not been demonstrated experimentally in an acoustic far-field platform. Second, the vast majority of implementations rely on tight-binding quantum models accessed via evanescent coupling and coupled-mode theory mappings to the Schrödinger equation~\cite{schrodinger_undulatory_1926}. While highly successful, this approach neglects collective far-field scattering effects, which are central to many wave systems and enable qualitatively different behavior. In particular, far-field coupling with equal linear inter-site hoppings sustains broadband propagation, opening the possibility of simultaneous conduction and topological confinement in the same spectral window — a dual-channel regime not accessible in tight-binding analogues with Bragg gaps. Although topological interface states have been demonstrated in far-field and locally resonant acoustic systems, these remain restricted to linear regimes~\cite{zhao_topological_2018,zangeneh-nejad_topological_2019,coutant_acoustic_2021,kaina_hermitian_2020,zhao_subwavelength_2021}. The identification and experimental realization of nonlinear topological phases in such scattering-dominated settings — combined with programmable, real-time active control — therefore remain largely open problems.

\begin{figure*}
    \centering
    \includegraphics[width=0.8\textwidth]{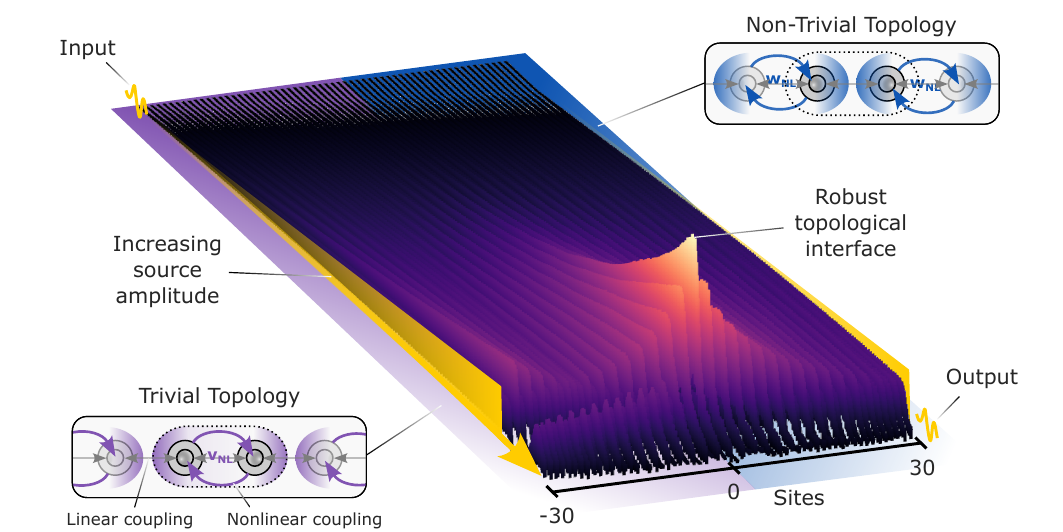}
    \caption{\textbf{Hybrid linear conduction and amplitude-driven topological confinement.} Numerical simulation of the scattering model of Sec.~II. Two finite chains (30 unit cells) with distinct topology form an interface at the centre. The insets specify the topological configuration of each side - trivial topology on the left (purple) or non-trivial topology on the right (blue). Grey arrows indicate linear coupling between the equally-spaced acoustic scatterers while the purple and blue arrows correspond to intra- and inter-cell couplings $v_{NL}$ and $w_{NL}$ respectively. The corresponding topological mode is input at the left of the chain for increasing source amplitudes. The equal-linear couplings lead to unimpeded conduction throughout the chain whereas the dimerizing nonlinear couplings cause the appearance of an amplitude-dependent topological interface mode at the center that is robust to disorder.}
    \label{fig:theory_1D}
\end{figure*}

Here, we address these challenges using a programmable chain of active electroacoustic resonators (AERs), enabling real-time electronic control of both local resonances and nonlocal couplings. By engineering nonlinear interactions through active feedback, we demonstrate the emergence of amplitude-dependent topological interface states coexisting with broadband linear transmission. As the excitation amplitude increases, the nonlinear dimerization grows continuously relative to the background linear coupling, progressively revealing a topological interface mode. Figure~\ref{fig:theory_1D} illustrates the effect in a simulation of the model developed below. Starting from a classical wave-scattering description, we derive an effective pseudo-Hamiltonian formulation analogous to quantum scattering theory, from which a minimal topological model naturally emerges. We show that this model supports chiral-symmetry-protected interface states with tunable amplitude dependence. These predictions are validated through circuit simulations and experiments on a fully programmable active acoustic metacrystal.

\section{Model and Hamiltonian description}
\begin{figure*}
    \centering
    \includegraphics[width=0.8\textwidth]{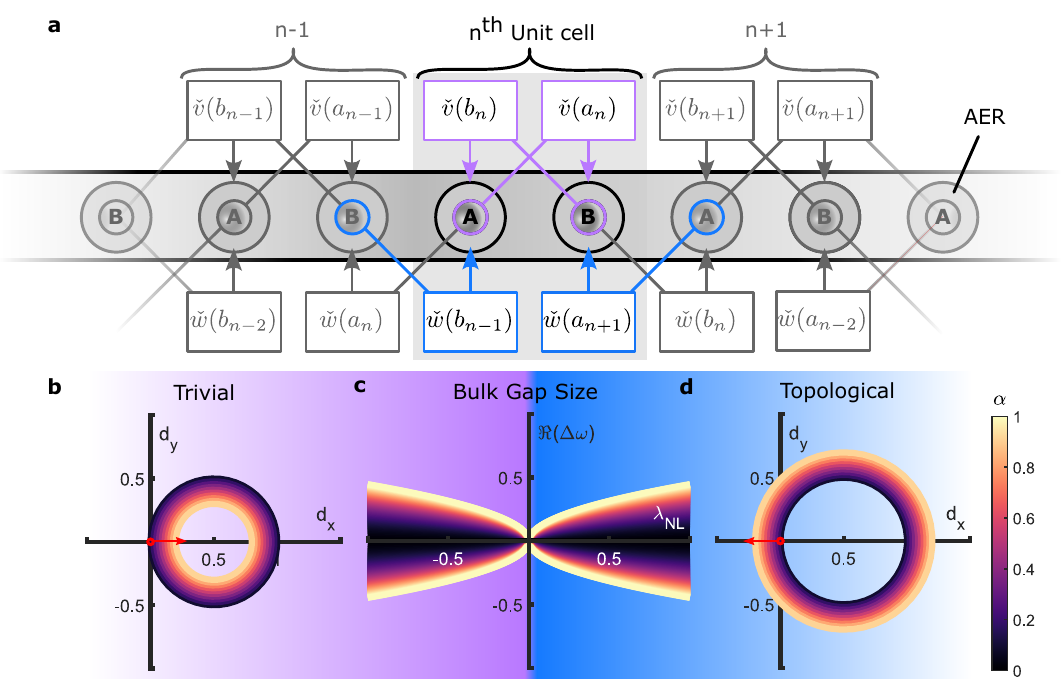}
    \caption{\textbf{Theoretical model for nonlinear topology.} \textbf{a} Coupling scheme for chain of nonlinearly dimerized scatterers. Two AERs labelled $A$ and $B$ constitute the unitcell. Intra- and inter-cell couplings (purple and blue arrows resp.) are conveyed by nonlinear amplitude-dependent parameters $\check{v}(p_n) := v_L + v_{NL}|p_n|^\nu$ and $\check{w}(p_n) := w_L + w_{NL}|p_n|^\nu$ where $p_n = a_n,b_n$ is the pressure on site $A,B$ of the n-th dimer respectively. \textbf{b-d} Amplitude-driven topological transition for equal linear coupling $v_L = w_L$ and where nonlinear couplings specify the topological regime. The critical point is at driving amplitude $\alpha = 0$: for any $\alpha>0$ the trajectory either encircles the origin (panel \textbf{d}) or does not (panel \textbf{b}), so what grows with $\alpha$ in panel \textbf{c} is the gap, and hence the confinement, not the invariant. We define the nonlinear dimerization parameter $\lambda_{NL} = w_{NL} - v_{NL}$ where $w_{NL} + v_{NL}  =1$. \textbf{b}~Trivial nonlinear coupling ($\lambda_{NL}<0$): Endpoint path of the d-vector representing the Hamiltonian, $\mathbf{H} = \mathbf{d}(q,\alpha)\cdot\mathbf{\sigma}$, for increasing dimer amplitude $\alpha$.  \textbf{c}~ Bulk gap size as a function of dimer amplitude $\alpha$ and the nonlinear dimerization parameter $\lambda_{NL}$. $\textbf{d}$~  Topological nonlinear coupling ($\lambda_{NL}>0$): Endpoint path of the d-vector for increasing dimer amplitude $\alpha$.}
    \label{fig:theory_topo}
\end{figure*}

In the single-mode regime, an acoustic scattering chain of dimers with linear (and nonlinear) \textit{intra}-cell and \textit{inter}-cell couplings, denoted $v_{(N)L}$ and $w_{(N)L}$, can be described by the state vector $\ket{\Psi_n} = [a_n, b_n]^T$, where $a_n$ and $b_n$ represent the complex pressure amplitudes at the $n^{\text{th}}$ unit cell on sublattices A and B, respectively (Fig.~\ref{fig:theory_topo}a). Its dynamics are governed by a discretized wave equation,

\begin{equation} \label{eq:scatter}
\begin{split}
(\mathbf{\Delta}_n + \mathbf{V}_n)\ket{\Psi_n}
+ \mathbf{W}_{n-1}\ket{\Psi_{n-1}}\\
+ \mathbf{W}_{n+1}\ket{\Psi_{n+1}}
=
\mathbf{\Delta}_t\ket{\Psi_n}
\end{split}
\end{equation}
where
\begin{gather*}
    \mathbf{\Delta}_n
    =
    \begin{bmatrix}
        \nabla_n^2  & 0 \\
        0 & \nabla_n^2
    \end{bmatrix},\quad
    \mathbf{\Delta}_t
    =
    \begin{bmatrix}
        c^{-2}\nabla_t^2  & 0 \\
        0 & c^{-2}\nabla_t^2
    \end{bmatrix}\\
    \mathbf{V}_n
    =
    \begin{bmatrix}
        0 & \check{v}(b_n) \\
        \check{v}(a_n) & 0
    \end{bmatrix},\quad
    \mathbf{W}_{n-1}
    =
    \begin{bmatrix}
        0 & \check{w}(b_{n-1})\\
        0 & 0
    \end{bmatrix},\\
    \mathbf{W}_{n+1}
     =
    \begin{bmatrix}
        0 & 0 \\
        \check{w}(a_{n+1}) & 0
    \end{bmatrix}\\
    \check{v}(x_n) := v_L + v_{NL}|x_n|^\nu ,\\
    \check{w}(x_n) := w_L + w_{NL}|x_n|^\nu ,\\  x_n \in [a_n,b_n]
\end{gather*}
(-- details are found in Supplementary Material I, ``Hamiltonian description of a classical metamaterial''~\cite{main_2_supp}).
For sufficiently weak nonlinearity parametrized by $\nu$, namely $v_{NL}\ll v_L$ and $w_{NL}\ll w_L$, the modal solution can be approximated to that of a time-harmonic Bloch wave $\ket{\Psi_{n}} = e^{i(q (n a) - \omega t)}\ket{\Psi_0}$ which, when substituted into Eq.\eqref{eq:scatter}, reveals a familiar eigenvalue problem:

\begin{equation}\label{eq:eigenval_prob}
    \mathbf{H}\ket{\Psi_0} = \left(\frac{\bar{\omega}}{c}\right)^2\ket{\Psi_0}
\end{equation}
where
\begin{multline}\label{eq:Hamiltonian}
    \mathbf{H}
    =
    \overbrace{
    \begin{bmatrix}
        0 &  v_L + w_L e^{+iqa}\\
        v_L + w_L e^{-iqa} & 0
    \end{bmatrix}}^{\mathbf{H_{L}}}\quad+ \\
    \underbrace{
    \begin{bmatrix}
        0 &  (v_{NL} + w_{NL} e^{+iqa})|b_{0}|^\nu\\
        (v_{NL} + w_{NL} e^{-iqa})|a_{0}|^\nu & 0
    \end{bmatrix}}_{\mathbf{H_{NL}}}
\end{multline}
and
\begin{equation*} \label{eq:E_omega_relation}
 \left(\frac{\bar{\omega}}{c}\right)^2 :=  \left(\frac{\omega}{c}\right)^2 - \frac{2}{a^2}
\end{equation*}
\\
Note that $|a_0|$ and $|b_0|$ correspond to the extended Bloch-wave amplitudes at sites $A$ and $B$ within the unit cell. We refer to $\mathbf{H}$ as a pseudo-Hamiltonian since its eigenvalues correspond to squared frequencies rather than energies. Composed of linearly and nonlinearly-coupled parts, $\mathbf{H}  =  \mathbf{H_{L}}+\mathbf{H_{NL}}$, $\mathbf{H}$ closely resembles that of a Schrodinger system from condensed matter or photonics. In fact, for $\mathbf{H_{NL}} = 0$, $\mathbf{H}$ directly maps to the Su-Schrieffer-Heeger model~\cite{su_solitons_1979}, albeit with wave-states corresponding to pressure modal amplitudes and the pseudo-eigenenergy that is quadratic in frequency. Indeed, first degree eigenvalues of a standard Hamiltonian fundamentally originate from the first order time differential operator in the Schrodinger equation whereas here, the system is described by the wave equation which is second order in time.
\section{Nonlinear topological model}\label{sec:NL-model}
We now aim to find a topological invariant for the two-part pseudo-Hamiltonian $\mathbf{H}  =  \mathbf{H_{L}}+\mathbf{H_{NL}}$. One strategy is to map $\mathbf{H}$ to a closed trajectory in parameter space and to define a topological invariant based on geometric arguments. Expressing $\mathbf{H}$ in terms of Pauli matrices would offer a promising pathway but the amplitude-dependent terms $|a_0|$ and $|b_0|$ in $\mathbf{H_{NL}}$ render it non-Hermitian and, consequently, non-Pauli-decomposable. The latter can be overcome by imposing equal linear couplings which leads to neighboring amplitudes being approximately equal: i.e. $v_L = w_L \Rightarrow |a_0| \approx |b_0|$ (-- see Supplementary Material II, ``$\mathbf{H}$ in terms of Pauli matrices''~\cite{main_2_supp}). Consequently, the pseudo-Hamiltonian $\mathbf{H}$ can completely be written in terms of Pauli matrices $\mathbf{\sigma}$:
\begin{equation}
    \begin{aligned}
    \mathbf{H} =  \underbrace{(d_{x,L}(q)+d_{y,NL}(q,\alpha))}_{d_{x}(q,\alpha)} \mathbf{\sigma_x} +\\
    \underbrace{(d_{y,L}(q)+d_{y,NL}(q,\alpha))}_{d_{y}(q,\alpha)}  \mathbf{\sigma_y}
    \end{aligned}
\end{equation}
where
\begin{equation}
\begin{aligned}
\begin{pmatrix}
d_{x,L}(q) \\
d_{y,L}(q)
\end{pmatrix}
&=
\begin{pmatrix}
v_L(1+\cos(q a)) \\
w_L \sin(q a)
\end{pmatrix},
\quad v_L = w_L \\[6pt]
\begin{pmatrix}
d_{x,NL}(q,\alpha) \\
d_{y,NL}(q,\alpha)
\end{pmatrix}
&=
\begin{pmatrix}
\left(v_{NL} + w_{NL} \cos(q a)\right)\alpha^\nu \\
w_{NL} \sin(q a)\alpha^\nu
\end{pmatrix}
\end{aligned}
\end{equation}
and $\alpha := (|a_{0}|^\nu + |b_{0}|^\nu)^{1/\nu}$ is the dimer mode amplitude. $\mathbf{H}$ is thus mapped to a circle in the $(d_x,d_y)$ plane and is equipped with a dimer mode amplitude-dependent radius. Note that the system is now partitioned into two distinct channels that operate in parallel: a permanent conducting channel assured by linear couplings and an amplitude-dependent insulating channel by nonlinear coupling. These channels correspond to independent contributions to the effective Hamiltonian rather than physically separable propagation paths.

The absence of a $\sigma_z$ component places the system in the chiral orthogonal class (AIII, or BDI once time-reversal symmetry is included) to which the Su-Schrieffer-Heeger model belongs, so that $W$ is quantized to $0$ or $1$. This quantization refers to the nonlinear channel: $\mathbf{H_{NL}}$ dimerizes, opens a gap and carries the invariant, while $\mathbf{H_{L}}$ with $v_L = w_L$ remains gapless and sustains propagation at all frequencies. Fig.~\ref{fig:theory_topo}b and Fig.~\ref{fig:theory_topo}d show amplitude-driven transitions to trivial or to topological regimes respectively for equal linear coupling $v_L=w_L$ depending on the nonlinear coupling configuration. The topological invariant is 1 if  $\mathbf{d}(q,\alpha)$ encircles the origin and 0 otherwise. Alternatively, it can be obtained directly via brute-force evaluation of the winding number $W = \frac{1}{2\pi i}\int_{-\pi}^{\pi}\frac{\dd}{\dd q}\ln (d_{x}(q,\alpha) - i d_{y}(q,\alpha) )\dd q$ \cite{zak_berrys_1989}. The pseudo-frequency band gap computed near the edges of the Brillouin zone, $\Delta \bar{\omega} = \pm c\sqrt{|(v_{NL} - w_{NL})\alpha^\nu/2|} $ (-- see Supplementary Material III, ``Dispersion and band gap calculation''~\cite{main_2_supp}), opens for non-equal nonlinear couplings and widens with increasing amplitude $\alpha$ as shown in Fig.~\ref{fig:theory_topo}c.

Finally, practical topological insulators are uniquely identified through their robustness to disorder or perturbations. Here, the system owes its topological protection to symmetric temporal coupling which translates to chiral symmetry of $\mathbf{H}$, similar to the SSH model \cite{su_solitons_1979}. Concretely, topological protection is lost if a phase shift $\Delta\omega t$ is introduced in the Bloch-wave solution,  $\ket{\Psi_0} = [a_{0},e^{i\Delta\omega t}b_{0}]^T$ which alters the eigenvalue problem \eqref{eq:eigenval_prob} to
 \begin{gather}\label{eq:chiral_symmetry_breaking}
    \begin{gathered}
     \begin{bmatrix}
         +d_z &  \check{v}(b_{0}) + \check{w}(b_{0})e^{+iqa}\\
         \check{v}(a_{0}) + \check{w}(a_{0})e^{-iqa} & -d_z
     \end{bmatrix}
     \ket{\Psi_0}
     = \\
     \left( \left(\frac{\bar{\omega}}{c}\right)^2 + d_z \right)\ket{\Psi_0}
     \end{gathered}
 \end{gather}
where $d_z = \frac{1}{2}(\Delta\omega/c)^2$. The new diagonal elements appearing in the pseudo-Hamiltonian correspond to a non-trivial Pauli-Z component, $d_z$,  and consequently break the topology-protecting chiral symmetry.

\section{Experimental realization}
\begin{figure*}
    \centering
    \includegraphics[width=0.8\textwidth]{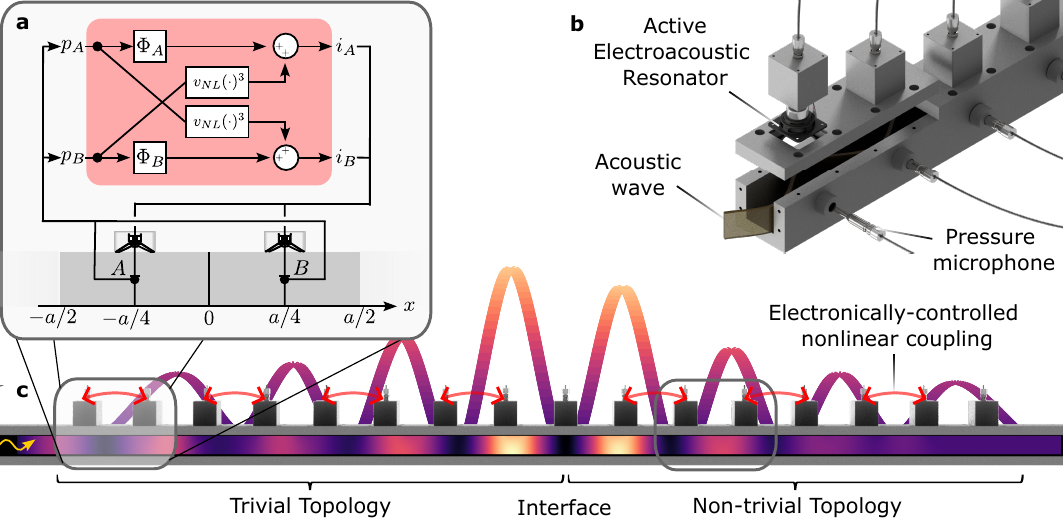}
    \caption{\textbf{Experimental apparatus and coupling scheme.} \textbf{a} Controlling nonlinear coupling in a unit cell. The pressures $p_{A,B}$ are probed at resonators $A$ and $B$ and are inputted and processed in a real time by a controller highlighted in red. From left to right, the pressures are read out and follow one of two paths: The first is denoted by the $\Phi_{A,B}$ operation which corresponds to local impedance synthesis of the resonator in view of reducing local losses (-- c.f. Supplementary Material IV~\cite{main_2_supp}). The second denoted by $v_{NL}(\cdot)^3$ corresponds to the controlled nonlinear coupling operation. Finally, electrical currents $i_{A,B}$ are outputted and conveyed to their respective resonators. \textbf{b} Segment of the active experimental duct highlighting the composition of a unit cell. Pressure microphones centered at the AERs serve both as real-time interaction control and mode probing (-- in phase and in amplitude). \textbf{c}  Cross-section of the full experimental metacrystal where the absolute pressure in the duct (also represented in the background) illustrates an interface mode driven at high amplitude excitations.  The electronically coupled pairs along the chain (red arrows) are configured such as to favor intra-cell coupling on one side and inter-cell coupling on the other.  The unit cell, indicated on both sides by the grey rectangles, highlight the different topological configurations. }
    \label{fig:setup}
\end{figure*}

\subsection{Design and configuration of the active metacrystal}
A finite metacrystal composed of 16 individually controlled AERs was developed and manufactured. The acoustic waveguide is composed of CNC-milled 20mm-thick PVC slabs lined with 16 off-the-shelf Visaton FRWS 5 - 8 Ohm drivers, 3D-printed back-cavities, and quarter-inch PCB microphones. The pressure microphones centered at the AERs serve both for interaction control and mode probing. Active control is carried out by a commercial Speedgoat machine equipped with the IO135 module. This Direct Memory Access (DMA) module enables both on-site tunability and low-latency driver control over all 16 channels. COMSOL Multiphysics finite element simulations were performed enabling design optimization and experimental validation for linear coupling configurations. In order to account for manufacturing inconsistencies, the mechanical mass, resistance and compliance of each speaker was characterized and actively tuned to match one another. Finally, the quality factor of the resonators were all enhanced by synthetically reducing AER acoustic resistance by 80$\%$ (-- details regarding this impedance synthesis are found in Supplementary Material IV~\cite{main_2_supp}).

A comprehensive overview of the experimental apparatus is presented detailing the electronic coupling scheme, the unit cell and the coupling configuration of the metacrystal (Figs.~\ref{fig:setup}a-c). The first half of the crystal is electronically configured so as to favor nonlinear intra-cell coupling, i.e. $v_{NL}>w_{NL}$, and the other to favor inter-cell coupling, $v_{NL}<w_{NL}$. In other words, the system is set to host a topological interface state at the center of the metacrystal. Equal linear couplings, $v_{L}=w_{L}$, are achieved without the need of active control as it is passively ensured by the equal spacing of the AERs along the duct. Conversely, nonlocal and nonlinear couplings are implemented electronically, with the nonlinear response arbitrarily tuned to third order via an active feedback loop, as illustrated in Fig.~\ref{fig:setup}a. In terms of the previously introduced topological model, this nonlinearity corresponds to $\nu = 2$. Note that the inter-site coupling strengths are chosen to be symmetric, but can be readily modified to achieve non-reciprocity, as demonstrated in our recent study~\cite{padlewski_observation_2025}.%
\subsection[Nonlinear topological interface state and broadband transmission]{Nonlinear topological interface state and broadband transmission}\label{sec:broadband}

\begin{figure}
    \centering
    \includegraphics[width=0.5\textwidth]{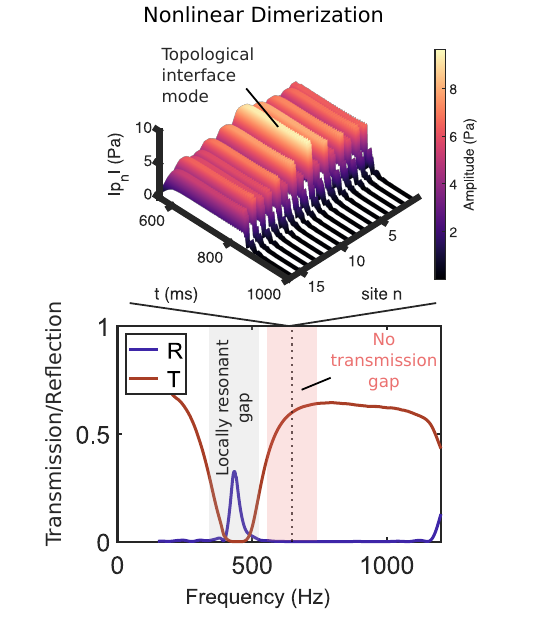}
    \caption{\textbf{Nonlinear topological interface state with broadband transmission.} Both panels are experimental measurements. \textbf{(Top)} Spatio-temporal evolution of the measured pressure amplitude $|p_n|$ across the 16 array sites when the metacrystal is driven at 635~Hz, showing a mode whose maximum sits at the interface site, superposed on the delocalized conduction channel sustained by the equal linear couplings. \textbf{(Bottom)} Measured transmission/reflection spectrum: a locally resonant gap appears at 480~Hz, the resonant frequency of the AERs, while transmission remains broadband throughout the Bragg-frequency range. See Supplementary Material V~\cite{main_2_supp} for the analogous linear SSH case, and Supplementary Material VI~\cite{main_2_supp} for the non-dimerized and uniformly trivially-dimerized control configurations.}
    \label{fig:exp_Smatrix_NL}
\end{figure}

The coexistence of broadband transmission and the nonlinear topological interface state are shown in Fig.~\ref{fig:exp_Smatrix_NL}. The transmission/reflection spectrum exhibits a locally resonant gap, associated with the resonant frequency of the AERs, but no Bragg gap: transmission remains broadband across the entire frequency range where the topological interface state would otherwise produce a spectral feature (see Supplementary Material V~\cite{main_2_supp} for comparison with the conventional linear SSH case). Independently, the spatio-temporal measurement in the same figure shows the topological interface mode with its maximum at the interface site when the metacrystal is excited at the Bragg frequency (see Supplementary Material VI~\cite{main_2_supp} for comparison with non-topological configurations). Together, these two measurements demonstrate that the topological mode persists in real space even though it leaves no signature in the linear transmission spectrum. This observation is consistent with the theoretical model, which predicts a broadband transmission mediated by the non-dimerizing linear couplings. Note that in the absence of the locally resonant elements, the measured transmission would be completely broadband.

\begin{figure*}
    \centering
    \includegraphics[width=1\textwidth]{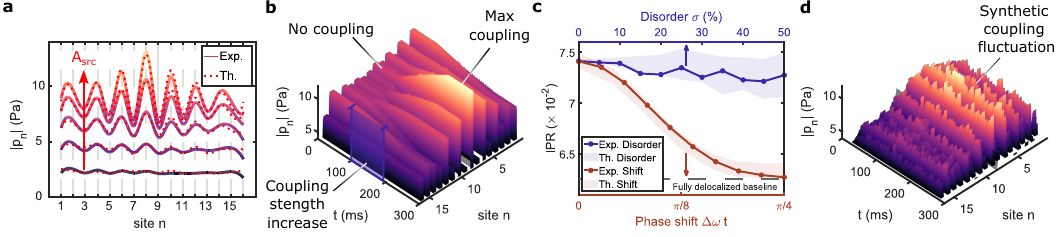}
    \caption{\textbf{Realization of robust amplitude driven topological interface via tunable active control.} All panels show experimental measurements; calculated dynamics are overlaid in \textbf{a} and \textbf{c}. \textbf{a} Amplitude-driven topological interface state: pressure at each site along the duct for five source amplitudes ($A_{\rm src} = 2.2$, $4.3$, $6.2$, $7.8$ and $9.0$~Pa), measurement (thick colored) and theory (dashed red line). \textbf{b} Real-time parameter tuning: spatio-temporal evolution of the pressure as the nonlinear coupling  is ramped adiabatically from zero in the region highlighted in blue, at a fixed source amplitude of 9~Pa. The sweep therefore begins at $w_{NL} = v_{NL}$ and traverses the topological phase boundary, where the nonlinear gap closes. \textbf{c} Topological robustness: inverse participation ratio as a function of coupling disorder $\sigma$ (blue) and of the phase shift $\Delta \omega t$ delaying the coupling responses of adjacent sites (red); shaded intervals span the extrema over 20 random coupling realizations. \textbf{d} Robustness to fluctuating disorder ($\langle\sigma \rangle_t=50\%$), analogous to lattice temperature in condensed matter systems.}
    \label{fig:exp_results}
\end{figure*}

\subsection{Amplitude-dependent interface modes}
Amplitude-dependent interface modes are experimentally realized, in close agreement with theoretical predictions. Figure~\ref{fig:exp_results}a compares the experimental and theoretical temporal dynamics of the topological mode for increasing source amplitudes $A_{\rm src}$, providing clear evidence of its growth with amplitude (for details regarding the theoretical temporal dynamics, see Supplementary Material VII~\cite{main_2_supp}). The nearly equal pressures measured at the metacrystal's extremities further confirm the presence of broadband conducting channels, with the minute observed transmission loss attributable to the mechanical resistance of the AERs and thermo-viscous dissipation within the waveguide.

While the effect of nonlinearity is demonstrated by increasing the source amplitude, we emphasize that it does \emph{not} arise from gain saturation. Instead, the nonlinearity is precisely defined via electronic feedback, and the localization of the interface state depends solely on the balance between the nonlinear coupling strength and the source amplitude. Importantly, the localization grows continuously with amplitude. This behavior reflects the pre-configured topological contrast between the two halves of the metacrystal: the trivial/nontrivial asymmetry is set before excitation, and increasing the source amplitude progressively strengthens the nonlinear dimerization relative to the background linear coupling — gradually revealing the interface mode. To further illustrate this programmability, the system can also be configured to enhance the nonlinear coupling at a fixed source amplitude (Fig.~\ref{fig:exp_results}b).
A quantitative measure of localization is provided by the inverse participation ratio (IPR), defined as $\mathrm{IPR} = \sum_{i=1}^{16}(|\bar{p}_i|^2)^2$ where $\bar{p}_i$ is the pressure normalized so that $\sum|\bar{p}_i|^2 = 1$. Its inverse $1/\mathrm{IPR}$ estimates the effective number of sites participating in the mode: a perfectly delocalized state over $N=16$ sites gives $\mathrm{IPR}=1/N\approx 6\times10^{-2}$, while increased localization raises it above this baseline. The observed rise from $6\times10^{-2}$ to approximately $7.5\times10^{-2}$ corresponds to the mode's effective spatial extent contracting from the full 16-site array to roughly 13 sites — a quantitative, reproducible signature of amplitude-driven localization (see Supplementary Material VIII~\cite{main_2_supp} for the full IPR vs.\ amplitude curve).  To be clear, this extent refers to the \emph{total} measured field, which superposes the interface mode with the delocalized conduction channel that the equal linear couplings sustain by design. Since that channel carries most of the energy, the IPR of the total field is pinned near its delocalized limit however tightly the interface component itself is confined. We therefore use the IPR as a \emph{relative} indicator --- through its monotonic rise with source amplitude (Fig.~\ref{fig:exp_results}a and Supplementary Material VIII~\cite{main_2_supp}), its insensitivity to coupling disorder, and its collapse under a coupling phase delay (Fig.~\ref{fig:exp_results}c) --- rather than as an absolute measure of confinement.
We now use this dynamic coupling control to reveal the underlying symmetry protecting the topological interface mode.

\subsection[Symmetry protection and the topological origin of the interface state]{Symmetry protection and the topological origin of the interface state}\label{sec:symmetry}

As evoked in Sec.~\ref{sec:NL-model}, one of the key phenomenological markers asserting a claimed topological state is its protection against system disorder. In multiple scattering systems, protection has been shown to be a direct consequence of symmetry constraints of the unit cell transfer matrix $M_{\mbox{cell}}$ - namely time-reversal symmetry and $M_{\mbox{cell}}^2 = 1$~\cite{zangeneh-nejad_topological_2019}. Here, as the scattering system has been expressed in terms of a pseudo-Hamiltonian $\mathbf{H}$, the symmetry protection is translated to the chiral symmetry of $\mathbf{H}$ i.e. $\sigma_z \mathbf{H} \sigma_z ^\dagger = -\mathbf{H}$, where $\sigma_z = [1,0;0,-1]$. The goal is then to demonstrate topological robustness against coupling disorder and that chiral protection, assured here by time-reversal symmetry, can be broken by introducing a phase shift between the coupled sites.

State robustness can be assessed by measuring localization as a function of system disorder. As defined in the previous section, the IPR quantifies spatial localization, with values ranging from the fully delocalized limit $\mathrm{IPR} = 1/N = 1/16$ to the fully localized limit $\mathrm{IPR} = 1$.

The coupling disorder strength $\sigma$ is defined as a random deviation of the couplings with respect to their nominal values.  The IPRs extracted from both theoretical and experimental dynamics (and averaged over 20 random coupling configurations) are plotted as a function of  $\sigma$ in Fig.~\ref{fig:exp_results}c (blue).

\begin{figure*}[!t]
    \centering
    \includegraphics[width=\textwidth]{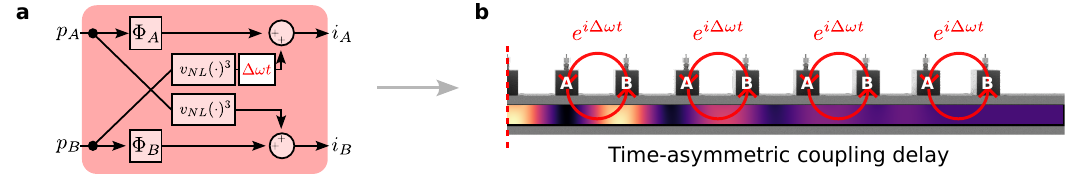}
    \caption{\textbf{Controlled breaking of the protecting chiral symmetry.} \textbf{a} Controller block diagram for a unit cell (\emph{cf.} Fig.~\ref{fig:setup}a), differing from it only by the tunable delay block $\Delta\omega t$ inserted after the nonlinear coupling operation feeding the current output $i_A$. The coupling magnitudes are left unchanged; only the relative timing of the two unidirectional couplings is affected. \textbf{b} Resulting time-asymmetric coupling along the array, shown on half of the metacrystal with a representative pressure distribution in the duct. The circular arrows mark the bidirectional coupling within each adjacent pair of resonators, to which the delay $e^{i\Delta\omega t}$ is applied, so that the coupling from site $A$ to site $B$ no longer acts simultaneously with that from $B$ to $A$. }
    \label{fig:chiral_break}
\end{figure*}

Time-reversal symmetry is preserved as long as coupling is simultaneous - i.e. the unidirectional non-local coupling from site $A$ to site $B$ occurs at the same time as the coupling from $B$ to $A$. This temporal symmetry can be trivially broken by introducing a delay or phase shift $\Delta\omega t$ between the unidirectional couplings of adjacent sites, implemented by inserting a tuneable time delay block after the nonlinear coupling operation feeding the current output $i_A$, as illustrated in Fig.~\ref{fig:chiral_break}. Again, the IPRs are extracted from both theoretical and experimental dynamics and plotted as a function of  $\Delta\omega t$ in Fig.~\ref{fig:exp_results}c (red).

The observed constant IPR as a function of coupling disorder highlights the system’s remarkable resilience to disorder. In fact, this robustness remains present for coupling disorders exceeding $50\%$ of their nominal values. Furthermore,  Fig.~\ref{fig:exp_results}d shows that the nonlinear topological interface state remains well defined even when real-time fluctuating disorder is introduced ($\langle\sigma \rangle_t=50\%$). Indeed, the disorder changes value after every feedback cycle and is reminiscent of lattice temperature in condensed matter.
Conversely, delaying the coupling response via $\Delta\omega t$  dramatically hinders the stability of the localized interface state as indicated by the rapid IPR drop with increasing shift. This phase shift effectively breaks time-reversal symmetry and,  as predicted by Eq.\eqref{eq:chiral_symmetry_breaking}, introduces a non-trivial $d_z$ component leaving the system topology ill-defined.

The two perturbations thus differ in symmetry rather than in severity: a $50\%$ randomization perturbs the coupling \emph{magnitudes} far more than the delay does, yet leaves the IPR unchanged, while the delay barely alters those magnitudes and destroys the state. Growth of the interface amplitude with drive would not by itself establish topology, since an amplitude-dependent coupling reshapes the propagating mode of any chain; a mode reshaped in that way would, however, show the opposite dependence here. This ordering --- together with the control configurations of Sec.~\ref{sec:broadband} --- therefore evidences the topological origin of the state independently of its absolute degree of localization.

\section{Conclusion}
In this work, we demonstrated an acoustic metamaterial exhibiting amplitude-driven topological confinement coexisting with broadband linear transmission across the same frequency range. This dual-channel behavior — simultaneous conducting and topological insulating response — arises from the orthogonal roles played by the linear and nonlinear inter-site couplings: the former sustains dispersionless propagation, while the latter introduce amplitude-dependent dimerization. The insulating response is a property of the nonlinear channel of the effective Hamiltonian, which is gapped and carries a quantized winding number, while the linear channel remains gapless and conducting at the same frequency; this is why the topological state leaves no signature in the measured transmission spectrum. This regime, which has no analogue in tight-binding systems relying on on-site nonlinearities or Bragg gaps, constitutes the central novelty of this work. For a system dimerized solely through nonlinear inter-site interactions, we derived an effective topological model directly from the classical wave equation, revealing chiral-symmetry-protected states with intrinsic amplitude dependence.

These predictions were validated through both temporal dynamics simulations and experiments on a fully programmable platform composed of individually controlled active electroacoustic resonators. Because the interface state carries no spectral signature, its topological origin was established by symmetry rather than by spectroscopy: the state survives randomization of the coupling magnitudes in excess of $50\%$, which preserves chiral symmetry, yet collapses under a delay between adjacent coupling responses, which breaks it, and it does not form at all in non-dimerized or uniformly trivially-dimerized configurations of the same array. This asymmetric response to two perturbations of comparable magnitude is what a chiral-symmetry-protected state predicts and what a mode reshaped by amplitude-dependent couplings does not.

Beyond their fundamental interest, these results open new avenues for wave manipulation in reconfigurable media. The dual-channel behaviour is of direct practical consequence: because confinement is carried by the nonlinear channel and requires no Bragg gap, the device redirects energy to a programmable location without sacrificing its transmission band, whereas gap-based topological devices necessarily forfeit the spectral window in which they operate. The result is an intensity-selective acoustic router with no moving parts, in which the interface position is set electronically and can be relocated during operation and in which the routed fraction is set by the drive level: at low amplitude the array transmits edge to edge, while at high amplitude, a growing share of the energy is held at the interface. The tolerance to $50\%$ coupling disorder documented in Fig.~\ref{fig:exp_results}c is equally relevant to implementation, since it corresponds to the component-to-component variation that any manufacturable array must absorb.

In general, the platform enables controlled exploration of nonlinear topological phases in scattering-dominated systems, as well as applications such as intensity-dependent waveguiding, analog emulation of relativistic wave equations, and the realization of time-modulated and nonreciprocal topological states. More broadly, this work establishes a versatile route toward dynamically programmable metamaterials operating beyond conventional tight-binding paradigms.
\\

\begin{acknowledgments}
This research was supported by the Swiss National Science Foundation under Grant No.~200020\_200498.
\end{acknowledgments}

\bibliographystyle{apsrev4-1}
\bibliography{main,main_2_supp}

\pdfximage{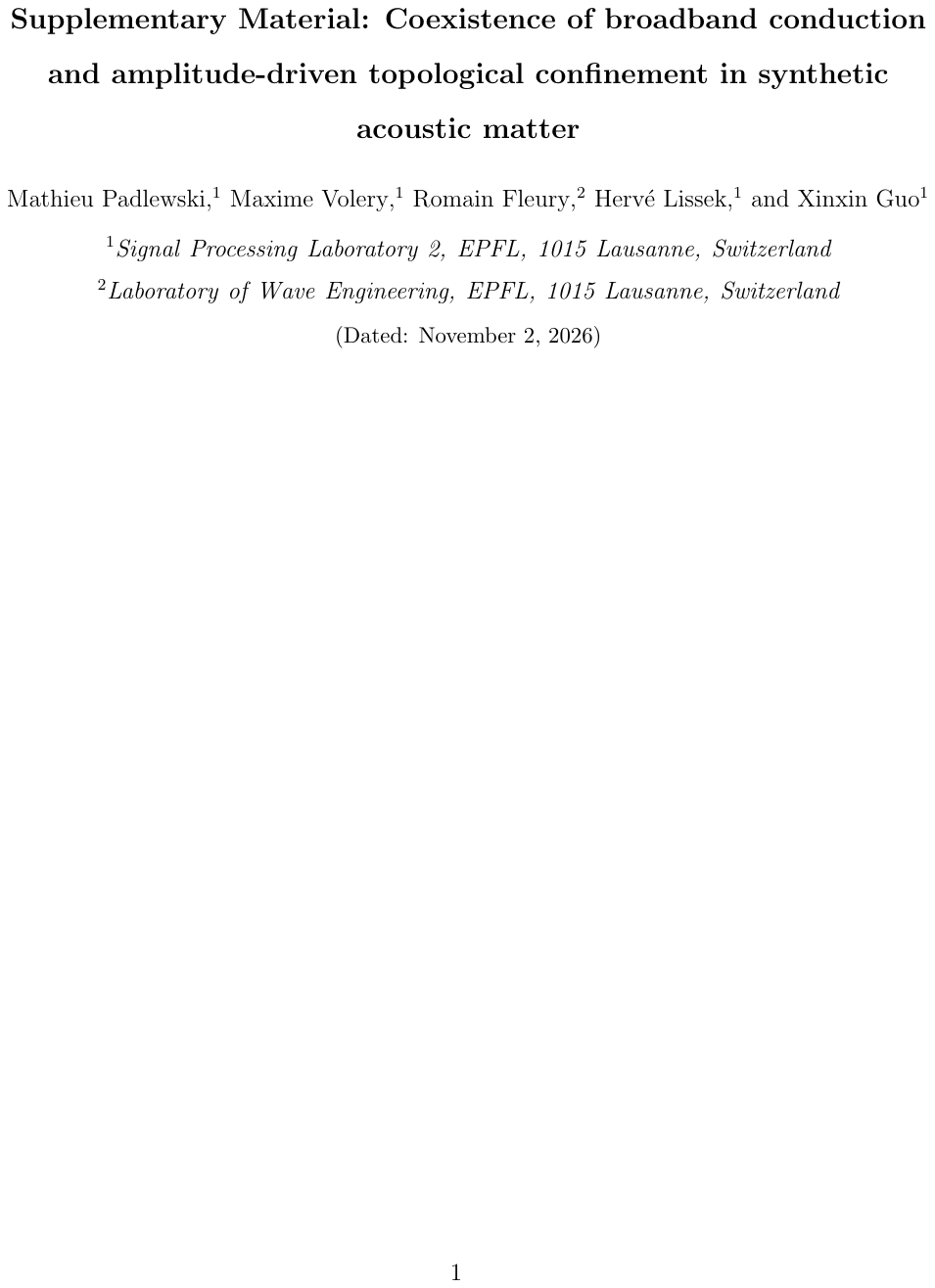}
\foreach \x in {1,...,\the\pdflastximagepages}{%
    \clearpage
    \includepdf[pages={\x},pagecommand={\thispagestyle{empty}}]{supp.pdf}
}

\end{document}